\documentclass{elsart}

\usepackage{epsfig}

\usepackage{amssymb}

\usepackage{amsmath}
\def\cmm2{{\,\rm cm^{-2}}}
\def\cm2{{\,{\rm cm}^2}}
\def\cmm3{{\,{\rm cm}^{-3}}}
\def\gcmm3{{\,{\rm g\,cm^{-3}}}}

\def\fun#1#2{\lower3.6pt\vbox{\baselineskip0pt\lineskip.9pt
  \ialign{$\mathsurround=0pt#1\hfil##\hfil$\crcr#2\crcr\sim\crcr}}}

\def\rhox{\,{\rho_{\rm X}}}
\def\px{\,{p_{\rm X}}}
\def\wx{\,{w_{\rm X}}}
\def\omegam{\,{\Omega_{\rm M}}}
\def\omegax{\,{\Omega_{\rm X}}}
\def\rhom{\,{\rho_{\rm M}}}

\begin{document}

\begin{frontmatter}


% use the thanksref command within \title, \author or \address for footnotes:
% \author{Name\thanksref{label2}}
% including your email address:
% \address{Address\thanksref{email}}
% \thanks[email]{E-mail: }

\title{Can luminosity distance measurements probe the equation of state of
dark energy}


\author{P. Astier}

\address{LPNHE, Universit\'es Paris VI \& VII, IN2P3/CNRS, 4 place Jussieu,
F-75252, Paris Cedex 05, France}

\begin{abstract}
% Text of abstract
Distance measurements to Type Ia supernovae (SNe Ia) at cosmological distances
indicate that the Universe is accelerating and that a large fraction
of the critical energy density exists in a component with
negative pressure. Various hypotheses on the nature of this 
``dark energy'' can be tested via their prediction for 
the equation of state of this component.
If the dark energy
is due to a scalar field, its equation of state will in general vary with
time and is related to the potential of the field.
We review the intrinsic degeneracies of luminosity distance measurements
and compute the expected accuracies that can be obtained for the equation
of state parameter 
from 
a realistic high statistic SNe Ia experiment.
\end{abstract}

\begin{keyword}
Cosmology \sep Dark Energy \sep Luminosity Distance \sep Type~Ia Supernovae 

\PACS 
\end{keyword}
\end{frontmatter}

% main text
%\section{}
%\label{}

%\begin{thebibliography}{00}

%\bibitem{}

%\end{thebibliography}

%\end{document}

\section{Introduction}

There is now strong evidence that the Universe is flat
and that matter only amounts to about 1/3 of the critical energy,
the remaining 2/3 exhibiting a large and negative pressure.
There are several candidates for this dark energy component,
which can be characterised by their ``equation of state'',
namely the ratio $w = p/\rho$. For example the genuine Cosmological 
Constant
has $p=-\rho$, while topological defects give
$p = -\rho/3$ for strings or $p = -2\rho/3$ for domain walls.
But dark energy could also be due to a possibly evolving scalar field,
in which case the equation of state may vary with time (and redshift).

Because they are performed at varying redshifts, luminosity distance 
measurements of Type Ia supernovae can, 
in principle,
provide
estimates of this possible varying equation of state.
The luminosity distance,
however, exhibits strong degeneracies, that may forbid any
precise determination of this equation of state, especially
if one allows it to vary with redshift. 

We first review
the expressions involved in the calculation of the luminosity distance,
and examine the main degeneracy that shows up when
analysing a simulated (yet realistic) high statistics SNe Ia experiment.
We then show how independent knowledge of $\omegam$ will limit
the effects of the degeneracy 
and compute the expected statistical and systematic 
uncertainties that can be achieved with such an experiment.

\section{Basic Equations}

We assume a flat universe made of 2 components~: 
non-relativistic matter, which contributes
$\omegam$ to the critical density,
and a single extra component $X$ described by 
its equation of state as a function of redshift:
\begin{equation}
\wx(z) = \px(z)/\rhox(z)
\end{equation}

The luminosity distance reads:
\begin{equation}
d_L(z) = (1+z) r(z) = (1+z) \int_{0}^{z} \frac{dz'}{H(z')}
\end{equation}
where r(z) is the Robertson-Walker comoving coordinate to an object
seen (via massless photons) at a redshift z.
Friedman's equation for a two component flat universe reads:
\begin{equation}
H^2(z) = H_0^2 ( \rhom(z) + \rhox(z))/\rho_0
\end{equation}
where $\rhom$ is the matter density, $\rhox$
is the dark energy density, and $\rho_0$ is the total density today.
The equation of state defines the way the component x
behaves with expansion:
\begin{equation}
\dot\rho_{\rm X} = -3H(1+\wx)\rhox
\end{equation}
Using $dz = -(1+z)H dt$, this equation can be integrated:
\begin{equation}
\label{eq:rhox_z}
\rhox(z) = \rhox(0) \exp{3 \int_0^z dz' \frac{1+\wx(z')}{1+z'}}.
\end{equation}
One may notice that $\wx = -1$ corresponds to a constant $\rhox$. 
For the matter component, $\rhom$, $w = 0$ 
which leads to $\rhom(z) = \rhom(0) (1+z)^3$.
Since observations favour a negative $\wx$ (\cite{42sn}), $\rhom/\rhox$ 
very likely increases with redshift.

The luminosity distance can be written:
\begin{equation}
\label{eq:dl_wxeq}
d_L(z) = 
\frac{1+z}{H_0} \int_{0}^{z} \frac{dz'}{\sqrt{\omegam(1+z')^3 + 
(1-\omegam)\rhox(z)/\rhox(0)}}
\end{equation}
$\wx(z)$ is the (unknown) function to be determined from measurements
of $d_L$ at various redshifts. For dark energy with negative pressure
($ \wx < 0 $i), 
the higher the redshift, the higher the relative matter contribution
in the above equation. 

To see how the unknown $\wx$ depends on the data $d_L$,
one may invert Equation~\ref{eq:dl_wxeq}, 
first by deriving it with respect to z:
\begin{equation}
\log {\frac{{r'}^{-2}- H_0^2 \omegam (1+z)^3}
 {H_0^2 (1-\omegam) }} = 
 3 \int_0^z \frac{1+\wx(z')}{1+z'},
\end{equation}
where $r(z) = d_L(z)/(1+z)$, and $r' = dr/dz$. Deriving once again yields:
\begin{equation}
\label{eq:wxfnr}
1+\wx(z) = \frac{1+z}{3}\frac{3 H_0^2 \omegam(1+z)^2 + 2r''/{r'}^3}
{H_0^2\omegam(1+z)^3-{r'}^{-2}}
\end{equation}
So, $\wx$ is related to the first and second derivatives of the 
distance in a highly non-linear way. Furthermore, the terms in the sums
composing the numerator and denominator of Equation~\ref{eq:wxfnr} are of 
comparable size and opposite sign: they typically differ by 30 \% 
at a redshift of 1. 
As a consequence,
small variations of the derivatives of r will result in large 
variations of $\wx$. This is shown the other way around 
in \cite{Maoretal}, through striking plots.

Several papers recently addressed the problem of reconstructing $\wx$ 
(or equivalently the potential of a scalar field) from luminosity 
distance measurements. Some conclude that it is hopeless given
expected uncertainties of large statistics SNe Ia measurements \cite{Maoretal}, 
others provide reasonably accurate estimations of $\wx(z)$, either based on
Monte-Carlo experiments \cite{Huterer_Turner_1}, or even on 
available SNe Ia data
\cite{Sainietal}. Before attempting to clarify what causes these fundamental 
differences, we will warn the reader about a potential misconception 
that may arise from Equation~\ref{eq:wxfnr}: both its numerator and denominator 
are insensitive to $H_0$ because r scales as $H_0^{-1}$. Hence w depends
on $\omegam$ and not on $\omegam H_0^2$, nor on $H_0$.

\section{Measurements and uncertainties}
Type Ia supernovae, have been used to constraint cosmological parameters
\cite{42sn,hizsn}. Using the light curve width-peak brightness empirical
relation,
the intrinsic dispersion of peak brightness of SNe Ia was found to be 
about 0.15 magnitude. A measurement with an total error of 0.2 magnitude
is already possible (it is the best resolution obtained in \cite{42sn}),
and we will assume conservatively this value as a standard error. It translates
into {\it relative} uncertainties of measured luminosity distances
of 10 \%, which makes the current redshift measurement uncertainty 
totally negligible.
For the proposed SNAP space mission \cite {snap_web}, 
the accessible redshift range is limited to 1.7, due
mainly to the overwhelming integration times required to reach
the expected signal to noise. There is also a limitation on the low
redshift side due to the large area that has to be covered 
to discover a sufficient number of 
low redshift supernovae.
%, which cannot be efficiently observed by a telescope 
%allowing to measure high redshift supernovae. 
Minimising systematic errors 
dictates that the whole sample is observed using the same apparatus.
We will therefore consider as a conservative baseline that 
2000 SNe Ia with redshifts in the range $[0.2,1.4]$ can be 
efficiently observed 
and see how well we can reconstruct $\wx(z)$.

If we have a dataset of SNe Ia with redshifts ${z_i, i=1..N}$,
we may compute least squares estimates of the cosmological parameters
by minimising:
\begin{equation}
\chi^2 = \sum_{i=1}^{N} \frac{(d_i-d_L(z_i,\theta))^2}{(p d_i)^2}
\end{equation}
where $d_i$ is the measured luminosity distance, $\theta$ is the set of
cosmological parameters to estimate, $d_L(z_i,\theta)$ is the expected
value of the luminosity distance for the redshift $z_i$ and the cosmological 
parameters $\theta$, p is the relative error (assumed to be 0.1) 
on the measurement. 

If one assumes that $\wx$ is constant($\wx = w_0$), 
$\theta$ is a 2 component vector $(\omegam,w_0)$. To estimate a varying $\wx$
without any prejudice, one may parametrise $\wx$ as a polynomial:
\begin{equation}
\wx = \sum_{i=0}^d w_i z^i
\end{equation}
Choosing a polynomial of z or 1+z does not make any difference from
the estimation point of view since there is a one-to-one 
mapping of the coefficients. 
Using this expression for $\wx$, 
Equation~\ref{eq:dl_wxeq} can be rewritten (for clarity we limit 
the polynomial to
d=2):
\begin{equation}
\label{eq:dl_wxp}
d_L(z) = 
\frac{1+z}{H_0} \int_{0}^{z} \frac{dz'}{\sqrt{\omegam(1+z')^3 + 
(1-\omegam) W(z') }}
%(1-\omegam)(1+z')^{w_0-w_1+w_2} \exp ((w_1-w_2)z'+w_2z^{'2}/2) }} 
\end{equation}
where $W(z)=(1+z)^{w_0-w_1+w_2} \exp ((w_1-w_2)z+w_2z^2/2)$.

For models with varying $\wx$,
we will concentrate on estimating the 3 parameters:
$(\omegam,w_0,w_1)$, which is enough to illustrate where the difficulties 
show up, for 3 different universes $(\omegam,w_0,w_1)$ = (0.3,-1.0,0.0),
(0.3,-0.8,0.3) and (0.4,-1.0,1.0), labelled A B and C in the figures.
Model A is the standard Cosmological Constant, model B is close to a 
quintessence model
with an inverse power law potential modified by 
supergravity~\cite{BraxMartin}, 
and model C is a toy model with a rapidly varying w.

The (logarithmic) derivatives shown in 
%Figure~\ref{fig:dlder3} 
Figure 1 determine 
the information that every supernova adds as a function of redshift.
It is clear that higher redshifts provide more information, but this 
becomes less clear if one considers that several lower redshift objects
can be measured in the time required for the measurement of one high 
redshift object.
One important point to notice is that for a given model,
all derivatives are very similar in shape which means that the parameter 
combination that a redshift probes does not vary strongly with redshift.
\begin{figure}[ht]
\centerline{\epsfig{figure=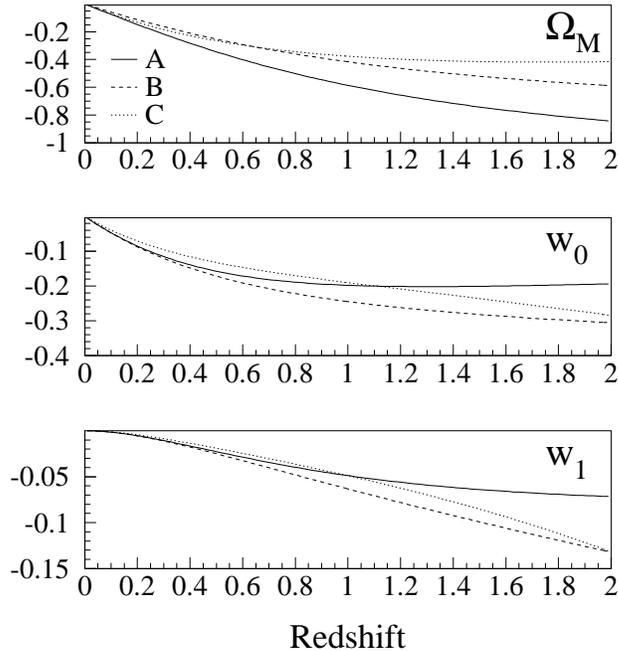 ,width=10cm}}
\caption{Logarithmic derivatives of the luminosity distance w.r.t 
$\omegam$, $w_0$ and $w_1$, for the 3 models under study (see text),
as a function of redshift.
One may notice similar shapes of the three derivatives for every model.
}
\label{fig:dlder3}
\end{figure}

To study what influences the variance (and covariance) of parameter estimates,
we will restrict ourselves to the quadratic approximation of the $\chi^2$
which consists in linearising $d_L$ as a function of parameters
around the chosen
model.
This is clearly not to be done for an actual estimation, where
an accurate mapping of the actual $\chi^2$ values (or of the likelihood)
is necessary to obtain reliable confidence contours. Within
this linear approximation, the Hessian of
the $\chi^2$ w.r.t the parameters reads:
\begin{equation}
\frac{1}{2}\frac{\partial^2 \chi^2}{\partial \theta^2} = F = 
   \sum_{i=1}^{N} h(z_i,\theta) h^T(z_i,\theta), \hspace*{0.5cm} 
   {\rm with} 
   \hspace*{0.5cm} 
    h(z_i,\theta) = \frac{1}{p d_L}\frac{d d_L(z_i,\theta)}{d \theta}
\end{equation}
All derivatives are usually evaluated at the minimum $\chi^2$ 
(i.e. the parameter estimate). As we are studying the estimation variance as a 
function of parameters and dataset, we will evaluate the derivatives at the parameter 
value, which should be the average estimate value.  
Since $\theta$ is vector, h is also a vector, and F is then a matrix, 
called the Fisher (or information or sometimes weight) matrix. 
Every $hh^T$ term of the sum is a matrix 
of rank 1 that is the information on the parameters that a given supernova adds.
Within this linear approximation, The covariance matrix of the estimates is 
just
$C = F^{-1}$.

\begin{table}[ht]
\begin{center}
\begin{tabular}{|c|c|c|c|}

\hline

model &  ``C'' & $\sqrt{\lambda}$ & Eigenvectors  \\
\hline
A &

 $ 
 \begin{matrix} 
0.2666  & -0.9853  & -0.9951  \\
-0.9853  & 0.2826  & 0.9643  \\
-0.9951  & 0.9643  & 2.117  \\
 \end{matrix}
 $ 

 &  

$ 
 \begin{matrix} 
0.00417 \\
0.0781 \\
2.152 \\
 \end{matrix}
 $ 
 
 & 

 $ 
 \begin{matrix} 
0.9418 & 0.3274 & 0.07586 \\
0.3126 & -0.9363 & 0.1598 \\
-0.1233 & 0.1268 & 0.9842 \\
 \end{matrix}
 $ 

% fin de A
\\
\hline
B &
$ 
 \begin{matrix} 
1.1913  & -0.9996  & -0.9976  \\
-0.999608  & 1.42776  & 0.9953  \\
-0.9976  & 0.9953  & 2.290  \\
 \end{matrix}
 $ 
 &  $ 
 \begin{matrix} 
0.0053 \\
0.125 \\
2.948 \\
 \end{matrix}
 $ 
 & 
 $ 
 \begin{matrix} 
0.8516 & 0.5086 & 0.1263 \\
0.3339 & -0.7123 & 0.6172 \\
-0.4039 & 0.4834 & 0.7765 \\
 \end{matrix}
 $
\\
% fin de B
\hline
C &
$ 
 \begin{matrix} 
0.2762  & -0.9964  & 0.6763  \\
-0.9964  & 0.5835  & -0.7339  \\
0.6763  & -0.7339  & 0.26285  \\
 \end{matrix}
 $ 
 &  $ 
 \begin{matrix} 
0.0060 \\
0.174 \\
0.675 \\
 \end{matrix}
 $ 
 & 
 $ 
 \begin{matrix} 
0.8840 & 0.4538 & 0.1111 \\
-0.2333 & 0.222 & 0.9465 \\
-0.4048 & 0.8627 & -0.3028 \\
 \end{matrix}
 $
\\
% fin de C
\hline
\end{tabular}
\caption{Covariance matrix of the estimates for our 3 models 
(where diagonal elements
are the rms errors and off-diagonal elements are the correlation coefficients). 
Second column gives the square root of eigenvalues, the third gives the 
corresponding
eigenvectors. One may notice that the last eigenvector
totally dominates the error budget. Especially for model B, the errors 
induced by the degeneracy forbid any independent measurement of the parameters.}
\end{center}
\label{tab:Cov}
\end{table}

%Table \ref{tab:Cov} 
Table 1 gives C and its eigenvalues and eigenvectors
within the linear approximation for our 3 universes, 
and for 2000 uniformly spread 
supernovae over the redshift range, the distance to each of them 
measured with a 
10 \% accuracy (0.2 mag): for model B (the worst case),
the ratio of major to minor axis
of the error ellipsoid is more than 500, which means that the parameter combination 
corresponding to the last eigenvector is almost unmeasured. 

The dataset as proposed does not allow to estimate {\it jointly} 
$\omegam$, $w_0$ and $w_1$ with a decent precision. 
However, one should notice that the problem is very different 
if $\omegam$ is frozen, 
in which case the errors on $w_0$ and $w_1$ go down by 
more than one order of magnitude. In \cite{Huterer_Turner_1,Sainietal},
$\omegam$ is fixed, 
%at variance with
unlike in \cite{Maoretal} where $\omegam$
is estimated and marginalized over. This explains partly why these papers 
reach 
different conclusions. 

The reason for this large degeneracy can be found by expanding in powers of z 
the expression for $H^2$ under the square root in Equation~\ref{eq:dl_wxp} 
(which is
probed observationally through $(dr/dz)^{-1/2}$):
\begin{equation}
\frac{H^2(z)}{H_0^2} = 1 + 3(1+\omegax w_0)z + 3/2(2+5 \omegax w_0 +  \omegax w_1+ 3 \omegax w_0^2) z^2 + o(z^3)
\end{equation}
where $\omegax = 1 -\omegam$.

Consider for a while $w_1 = 0$: at first order in z, only $\omegax w_0$ 
is determined 
(as shown by the shape of the confidence contours in the $(\omegam,w_0)$ 
plane shown in \cite{42sn}), the second order fixes $w_0$ 
through $\omegax w_0^2$,
which is rather weak. $(\omegax w_0, \omegax w_0^2)$ are much closer 
to observable
that $(\omegam, w_0)$, but $\omegax w_0^2$ really lacks physical sense. 
The scheme is the same for $w_1$: it enters through $\omegax w_1$ 
at second and third order and through $\omegax w_0 w_1$ at third order.
As a consequence
the parametrisation $(\omegam, \omegax w_0, \omegax w_1)$ is not much less
degenerate than the one considered here.

Obviously, the main degeneracy goes away when fixing $\omegam$. Before 
studying how the uncertainties on $\omegam$ affect w, we will evaluate
the effect of optimising the redshift distribution of the dataset. 
%how better a dataset with an optimised redshift distribution is.

\section{Optimising the redshift distribution of the dataset}
To optimise the redshift distribution of the dataset, 
we first study the effect on the largest eigenvalue of C,
of adding to the total sample a small number of supernovae (e.g. 40)
in a given redshits bin.
%Figure~\ref{fig:plotv3} 
Figure 2 shows how this quantity 
evolves as a function of the redshift in which the 40 supernovae
were added.
%if one adds 40 supernovae as a function
%of the redshift of this added subsample. 
Although no dramatic change is seen 
(errors 
scale as the plotted values), 
%Figure~\ref{fig:plotv3} 
Figure 2 shows that 
a small subsample collected at much higher redshift would reduce 
the errors 
more
efficiently than collecting more events in the original redshift interval.

\begin{figure}[ht]
\centerline{\epsfig{figure=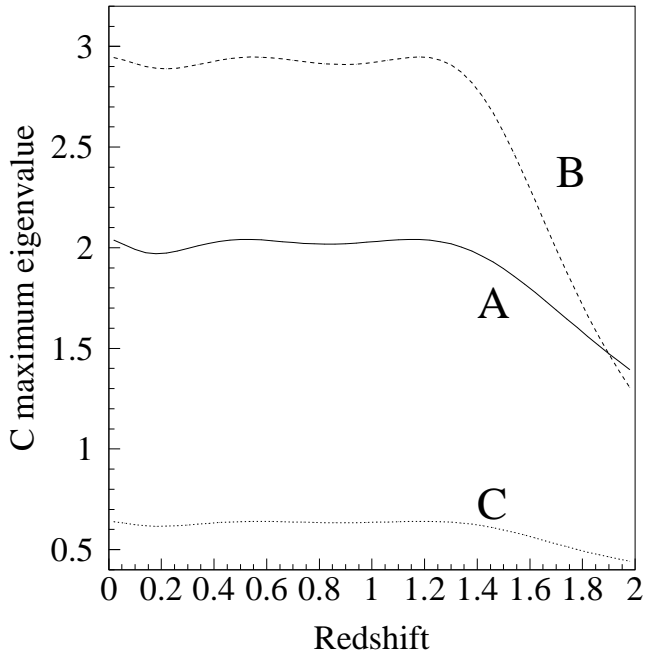 ,width=10cm}}
\caption{value of the square root of the largest eigenvalue of C as a function 
of the redshift where one adds 40 supernovae to the sample.}
\label{fig:plotv3}
\end{figure}
One may want to be more radical, i.e. really optimise the redshift 
distribution within
the redshift range in order to minimise this largest eigenvalue, 
keeping constant the
total number of supernovae. This does not yield
the same result as minimising the determinant of C as done 
in \cite{Huterer_Turner_2} where 
the optimal determinant is obtained at the cost of an even higher
largest eigenvalue. However, as for the minimum determinant, the 
optimum 
is reached for a redshift distribution 
consisting in 3 delta functions, that are given in 
%Table~\ref{tab:three_redshifts}.
Table 2.
The improvement with respect to a flat distribution is not large 
enough to consider
seriously such an extreme option, which has very severe drawbacks.  
Nevertheless,       
a moderate optimisation of the redshift distribution 
would be worth doing since it would lead to a gain in resolution.
%in resolution.
%, which is worth being considered.

\begin{table}[ht]
\begin{center}
\begin{tabular}{|c|c|c|c|}

\hline

model &  ``C'' &  z1 z2 z3  & fractions \\
\hline
A &
$ 
 \begin{matrix} 
0.1714  & -0.9720  & -0.989  \\
-0.972  & 0.1941  & 0.9295  \\
-0.989  & 0.9295  & 1.341  \\
 \end{matrix}
 $ 
 &
$
\begin{matrix} 
0.20 \\
0.82 \\
1.40 \\
 \end{matrix}
$
&
$
\begin{matrix} 
0.40 \\
0.40 \\
0.20 \\
\end{matrix}
$
\\
%fin de A
\hline 
B &
$ 
 \begin{matrix} 
0.7798  & -0.9989  & -0.9944  \\
-0.9989  & 0.9393  & 0.9887  \\
-0.9944  & 0.9887  & 1.478 \\
\end{matrix}
 $ 
 & 
 $ 
 \begin{matrix} 
0.20 \\
0.91 \\
1.40 \\
\end{matrix}
 $ 
 & 
 $ 
 \begin{matrix} 
0.34 \\
0.41\\
0.25\\
 \end{matrix}
$
% fin de B
\\
\hline 
C &
$ 
 \begin{matrix} 
0.1828 & -0.9918  & 0.3483  \\
-0.9918  & 0.3715  & -0.4583  \\
0.3483  & -0.4583  & 0.1957  \\
 \end{matrix}
 $ 
 & 
 $ 
 \begin{matrix} 
0.20 \\
0.86 \\
1.40 \\
\end{matrix}
 $ 
 & 
 $ 
 \begin{matrix} 
0.50 \\
0.36\\
0.14\\
 \end{matrix}
$
% fin de C
\\
\hline 

\end{tabular}
\caption{Result of optimisation of the redshift distribution to reduce 
the largest eigenvalue of C (where C is given the same way as in 
%Table~\ref{tab:Cov}).
Table 1).
The 3 optimal distributions are rather similar.}
\end{center}
\label{tab:three_redshifts}
\end{table}

\section{Expected accuracy on $w_0$ and $w_1$ imposing an $\omegam$ prior}
We computed in the Bayesian approach the $w_0$,$w_1$ joint confidence 
contours,
by marginalizing the estimate probability distribution over $\omegam$
with a Gaussian prior, without recourse to the linear approximation, 
%(see Figure 3).
%Figure~\ref{fig:w0w1contours}. 
We used a Gaussian prior with $\sigma = 0.05$,
which is 2 to 3 times better than nowadays precision from large scale 
structures,
and other methods.
The results are shown in Figure 2.
Despite our conservative inputs, model A and model B 
can be separated at more than 95 \% CL, whatever is the true one. 
Model B can be separated from domain walls ($(w_0,w_1) = (-2/3,0)$) 
with the 
same accuracy. 
\begin{figure}[ht]
\centerline{\epsfig{figure=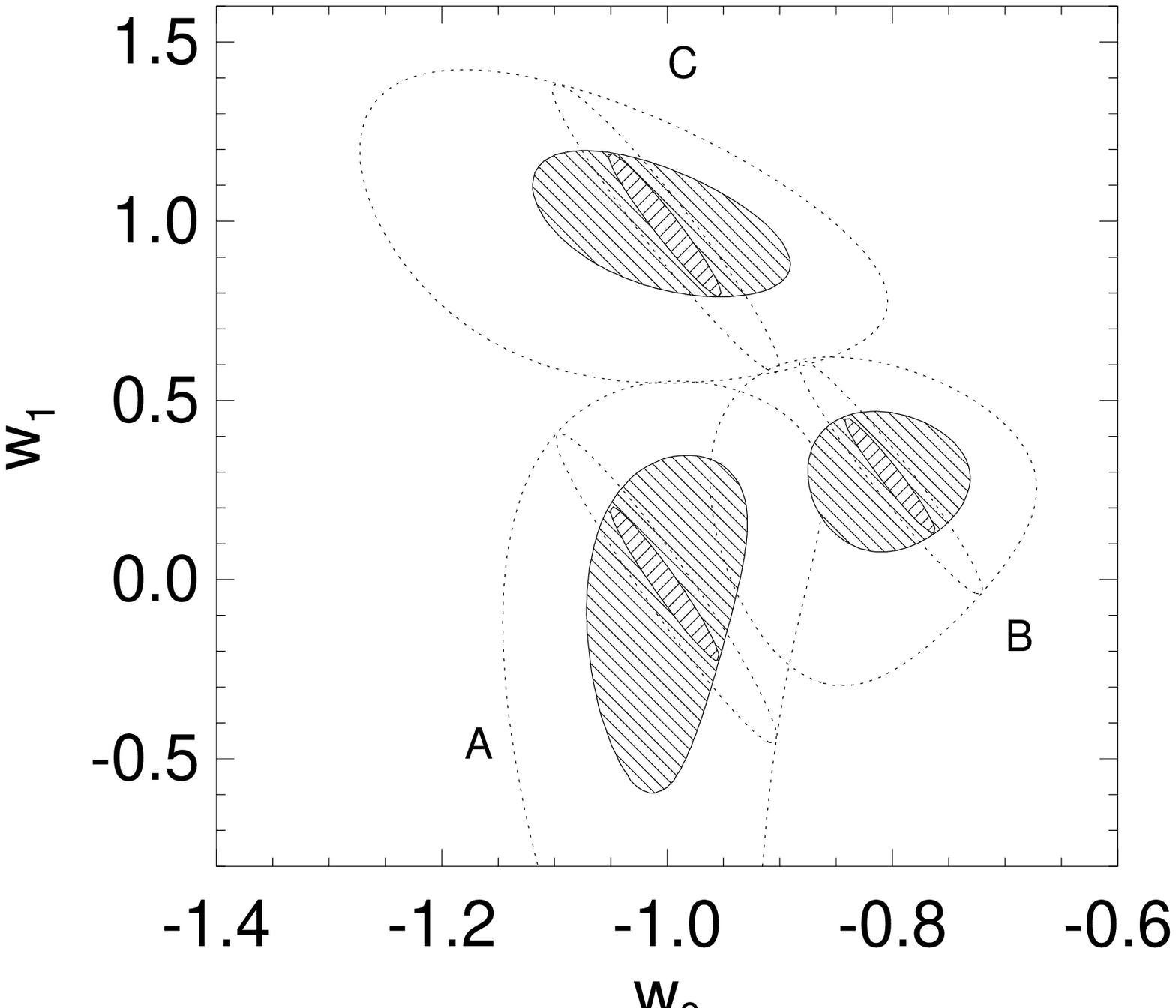 ,width=11cm}}
\caption{1 and 2 $\sigma$ (39\% filled areas,86\% dotted lines) 
confidence contours 
for ($w_0$,$w_1$) with a Gaussian prior
$\sigma_{\omegam} = 0.05$ (large contours), and with an infinite precision 
(small elongated contours). For model A, the 2 $\sigma$ contour escapes 
in an unphysical region.}
\label{fig:w0w1contours}
\end{figure}

It is interesting to come back again to the linear approximation
to study how estimated errors scale with the (external) $\omegam$
uncertainty, and the (internal) precision of the luminosity distance 
measurements.
Adding a prior is adding a $((\omegam-\omegam')/\sigma)^2$ to the $\chi^2$,
where $\omegam'$ is the prior value, and $\omegam$ remains the parameter 
to be estimated.
The Fisher and covariance matrix become:
\begin{equation}
F' = F + \frac{pp^T}{\sigma^2}, \hspace*{1cm}  
C' = C - \frac{(Cp)(Cp)^T}{\sigma^2 + p^T C p}
\end{equation}
where $C = F^{-1}$ and $p=(1,0,0)$ is the (unit) vector describing 
the prior (here it only concerns $\omegam$) and $\sigma$ is its standard error. 
The a posteriori variance of e.g. $w_1$ (after application of the prior) reads:
\begin{eqnarray}
V(w_1)& = & C'_{33} = C_{33} - \frac{(C_{31})^2}{\sigma^2+C_{11}} =
C_{33} ( 1- \frac{\rho_{13}^2}{\sigma^2/C_{11}+1}) \nonumber \\
      & \simeq &
C_{33}  (1 - \rho^{2}_{13} +  \rho^{2}_{13}\sigma^2/C_{11})
\end{eqnarray}
where the last approximation holds for $\sigma^2 << C_{11}$ (which is already
the case with nowadays precision on $\omegam$). $V(w_1)$ scales with $\sigma^2$
for large $\sigma$, and with $C_{33}$ for smaller ones. When 
$\sigma = C_{11} (1-\rho_{13}^2)/\rho_{13}^2$, $V(w_1)$ is twice 
its minimum value ($\sigma = 0$).
This happens for $\sigma = 0.03, 0.08, 0.44$  for models A,B and C 
respectively. Doing the same exercise for $w_0$ yields 
$\sigma = 0.05, 0.03, 0.02$.
With $\sigma = 0.05$ and our assumptions for the experimental precision, the
result quality would benefit both from a smaller $\sigma$ and from a better 
luminosity distance measurement.

For an infinitely well defined $\omegam$, we face a second ``degeneracy'' 
demonstrated
by the elongated shape of the error contours. The orientation of the 
major axis
defines a z ($z_{min}$) for which $Var(w_0+w_1z)$ is minimal, 
$z_{min} \simeq$ 0.2
to 0.3. Models with
similar $\wx(z_{min})$ but different $w_1$ (within $\sim$ 0.3 in 
our assumptions),
will be indistinguishable. This is shown by Figure~1 of 
\cite{Maoretal} where luminosity
distances for different $w_X$ having similar $w_X(z=0.3)$ are shown 
to have the same
luminosity distance behaviour.

\section{Influence of systematic errors}
%Systematic errors cause measured luminosity distances to be on average 
%different
%from true ones.
The overall scale of luminosity distances (deduced from fluxes)
depends on $LH_0^2$, where L is the absolute luminosity of the candle.
So distortions that bias the cosmological parameters are the ones that 
distort the shape
of $d_L$. At lowest order in z, we may model:
\begin{equation}
d_L^{measured} = d_L^{true} (1 +\alpha z), \hspace{1cm}
\end{equation}
where $\alpha$ (hopefully small) accounts for example for 
the drift of photometric calibration of supernovae across the redshift range,
an unknown evolution of supernovae, or light absorption in the intergalactic 
medium
uniformly over the spectral range. The SNAP mission has been designed to 
reduce
these effects as much as possible, and targets a precision of 0.02 mag 
over the 
redshift range $z<1.7$ which corresponds to $|\alpha| < 0.006$. 

We computed the parameter biases for our 3 models,
with $\sigma_{\omegam} = 0.05$ and $\alpha = 0.01$, and found negligible 
shifts 
for $\omegam$ (as expected), and biases for $(w_0,w_1)$ of (0,-0.19), 
(-0.01, 0.10) and (-0.02,-0.12) for models A,B and C. 
These systematic biases are all below the statistical uncertainty, 
and do not change 
significantly
for  $\sigma_{\omegam}$ = 0.1 or 0.025. 

\section{Conclusions}

With conservative hypotheses for the expected performances of a 
future high statistics SNe~Ia experiment such as the proposed SNAP mission, 
we find that the equation of state parameter of the dark energy 
can be measured 
provided an accurate value of $\omegam$ is used as a prior. 
%
%Despite conservative hypotheses concerning the plausible performance
%of the SNAP experiment for luminosity distance measurements, we reach
%the conclusion that the science goal may be met provided an accurate value
%of $\omegam$ is measured by other means. 
%One promising way to improve 
%the accuracy is detailed srudy of Type~Ia
%the ``standardity'' of SNe Ia is the planned supernova factory\cite{SNF}.
Detailed study of Type~Ia supernovae such as with  
the planned nearby supernova factory project~\cite{SNF} could result in 
further reducing the peak-brightness intrinsic dispersion which 
directly affects the precision on the equation of
state parameter.
In addition, independent
gain of precision can be achieved by 
(moderately) optimising the redshift distribution of the dataset.
%
%One may totally independently gain some resolution on the cosmological 
%parameters
%by (moderately) optimising the redshift distribution of the dataset.
Finally, it is interesting to note that
the required precision on $\omegam$ could come from the SNAP 
mission
itself, using large scale weak lensing which is sensitive to $\omegam$ 
almost 
independently of dark energy. In \cite{VanVerbekeEtAl}, a ground based 
moderately deep 
weak lensing survey of 5x5 degrees is estimated to measure $\omegam$ to 
10\% (for $\omegam$ around 0.3), reaching 5\% for a 10x10 degrees survey.

\ack
%\vspace*{5mm}
%{\it Acknowledgements.} 
It is a pleasure to acknowledge the stimulating 
discussions
among the FROGS (FRench Observing Group of Supernovae), especially with
R. Pain and J.M. Levy who also reviewed critically this manuscript.

\end{document}